\documentclass[letterpaper]{article} 
\usepackage{aaai2026}
\usepackage{times}  
\usepackage{helvet}  
\usepackage{courier}  
\usepackage[hyphens]{url}  
\usepackage{graphicx} 
\usepackage{natbib}  
\usepackage{caption} 
\usepackage{algorithm}
\usepackage{algorithmic}
\usepackage{subcaption} 
\usepackage{amsfonts} 
\usepackage{newfloat}
\usepackage{listings}
\DeclareCaptionStyle{ruled}{labelfont=normalfont,labelsep=colon,strut=off} 
\floatstyle{ruled}
\newfloat{listing}{tb}{lst}{}
\floatname{listing}{Listing}
\title{i-EXAM: Instructable and Explainable Attack Connectivity Graph Modeler}

\author {
    Rakesh Podder,
    Wadia Ganim,
    Sarath Sreedharan, 
    Indrajit Ray,
    Indrakshi Ray
}
\affiliations {
    Colorado State University Fort Collins, Colorado, USA \\
    \{Rakesh.Podder, Wadia.Ganim, Sarath.Sreedharan, Indrajit.Ray, Indrakshi.Ray\}@colostate.edu
}

\usepackage{bibentry}

\begin{document}

\maketitle

\begin{abstract}

i-EXAM is a planning-powered tool that helps system administrators to  create security profiles of complex networks and perform what-if analyses to identify network hardening strategies.
It leverages planning compilation
that provides soundness and completeness guarantees 
— to identify attack paths, evaluate security metrics, generate diverse hardening strategies, and explain these strategies in natural language using Large Language Models. 

\textbf{Demo Video}: https://youtu.be/c4m6EDIpMxM
\end{abstract}


\section{Introduction}
Securing and administering large networks is challenging:
 a single modification to a component in the network can have an enormous impact on its overall security posture and resiliency.
The SPEAR framework proposes how the Attack Connectivity Graph, (ACG) \cite{podder2025spear} 
a hypergraph, can be  compiled into a PDDL planning problem and help system administrators (sysadmins) identify the impact of an 
attack on the security and functionality of the network. 

We now present i-EXAM ({\em instructable and EXplainable Attack connectivity graph Modeler}), an interactive tool, built on the SPEAR framework, that leverages state-of-the-art AI planners  for network hardening \cite{podder2025spear}.  
i-EXAM, allows sysadmins to (a) visualize the overall network architecture, (b) run analysis methods to identify the overall security posture, (c) highlight potential attack paths, (d) identify diverse ways to update the network to achieve desired level of functionality or security hardening goals,  and (e) generate explanations for the sysadmin as to how the proposed changes would achieve the specified goals. 
i-EXAM can generate the planning models 
directly from network data and configuration files using a  suite of different planners and 
enables reasoning about attack paths and defensive strategies with formal guarantees. 
i-EXAM uses LLMs that interact with the planner to provide  explanations in
natural language for the sysadmins. 
\section{i-EXAM Tool}

\noindent{\bf Building Planning Models}
i-EXAM collects detailed system/host-level attributes such as software, version, files, etc. \& network-level data (i.e., port, protocol, IP, router, switch) using Nmap, Wazuh, and from documented data. It utilizes OpenVAS, Wazuh, and other scanners to get up-to-date vulnerability (CVE) information from external databases like NVD, ExploitDB. The information is then represented in JSON format, and used to create a database. This database is used as input to a PDDL generator to automatically build planning models.

\noindent{\bf Node/Path Visualizations}
i-EXAM has a visualization system, designed to provide sysadmins with
views of the security posture at different levels of abstraction.
The network information from the database is loaded to render the entire network as a graph, where each node represents a host and an associated vulnerability, and an edge represents network connections between the hosts.
A host having multiple vulnerabilities is represented as multiple nodes, thus providing a better visualization of attack paths.
Clicking on the node allows sysadmins to see additional information like the vulnerability ID, vulnerability severity scores,
 including Common Vulnerability Scoring System (CVSS) score, software, versions, file path, capabilities, cause, vulnerability types, and functionalities.
i-EXAM allows visualization of potential paths, which are calculated from the planning model and then overlaid on this visualization.
It uses a top-k-planner \cite{katz-et-al-icaps2018} to visualize multiple attack paths to the same target node.


\noindent{\bf Network-Level Analysis}
While i-EXAM can support any analysis that can be performed over an ACG, we focus on two metrics \cite{podder2025spear}. The first, the \textit{impenetrability metric} (M1), is defined as $\mathcal{F}^I_A(\mathbb{E}_A) = 1$ iff $|\mathbb{E}_A| = 0$, where $\mathbb{E}_A$ is the set of valid attack paths --- a binary indicator of whether any valid attack path exists. Its planning compilation directly encodes the ACG into a planning model $\mathcal{M}^G$ where fluents represent host-attribute pairs, actions represent attack/connectivity edges, and the target attribute becomes the goal; by Theorem 1~\cite{podder2025spear}, a plan exists iff an attack path exists. The second, the \textit{attack difficulty metric} (M2), captures the minimum-cost attack path: $\mathcal{F}^D_A(\mathbb{E}_A) = \min\{|E| \mid E \in \mathbb{E}_A\}$. Since M2 requires minimizing over multiple attacker initial positions and targets simultaneously, its compilation introduces auxiliary fluents (\texttt{init\_change\_mode}, \texttt{act\_mode}, \texttt{goal\_reached}) that allow the planner to select the cheapest attacker scenario, with optimal plan cost equal to $\mathcal{F}^D_A$ (Proposition 4,~\cite{podder2025spear}). Both metrics enforce $\mathcal{F}^1_C$, ensuring hardening strategies preserve service connectivity. These compiled problems are solved using $A^*$ with the LMCut heuristic~\cite{helmert2009landmarks} via FastDownward~\cite{helmert2006fast}.

\noindent{\bf What-If-Analysis}
 Security posture improvement consists of a set of modifications to the network 
 that is designed to reduce the metric values, while ensuring that the changes don't affect the availability of  hosted services. 
Note that, ACG, instead of more traditional attack graphs \cite{ou-mulval-2005}, allows us to simultaneously reason about both attacks and functionalities. 
Following the formulation set in previous works (cf. \cite{srivastava2007domain}), i-EXAM leverages a diverse solution selection, which will then be shown to the sysadmin. 
The diversity ensures that each set of modifications focuses on different components of the network. 
This allows i-EXAM to account for the fact that the cost of a modification is quite hard to capture exactly, and may depend on a number of factors, including the expertise of the sysadmin and the availability of technicians.
By providing sysadmins multiple diverse solutions, they will be in a position to make an
informed choice.

\noindent{\bf Explanation Generation}
i-EXAM also provides an explanation as to why the proposed modifications will help reduce the target metric. For this, i-EXAM leverages a modified version of explanations by model restriction \cite{krarup2021contrastive}, wherein it presents multiple counterfactual plans that would have satisfied previous metric value and how they are invalidated by the proposed changes. i-EXAM uses an LLM, specifically llama-3.1-nemotron-70b-instruct \cite{dubey2024llama}, to convert the plan failure information into natural language. This allows the tool to be easily used by sysadmin who may not have expertise on automated planning systems.

\section{Instantiation of i-EXAM}

\begin{figure}[h]
    \centering
    \includegraphics[width=\linewidth]{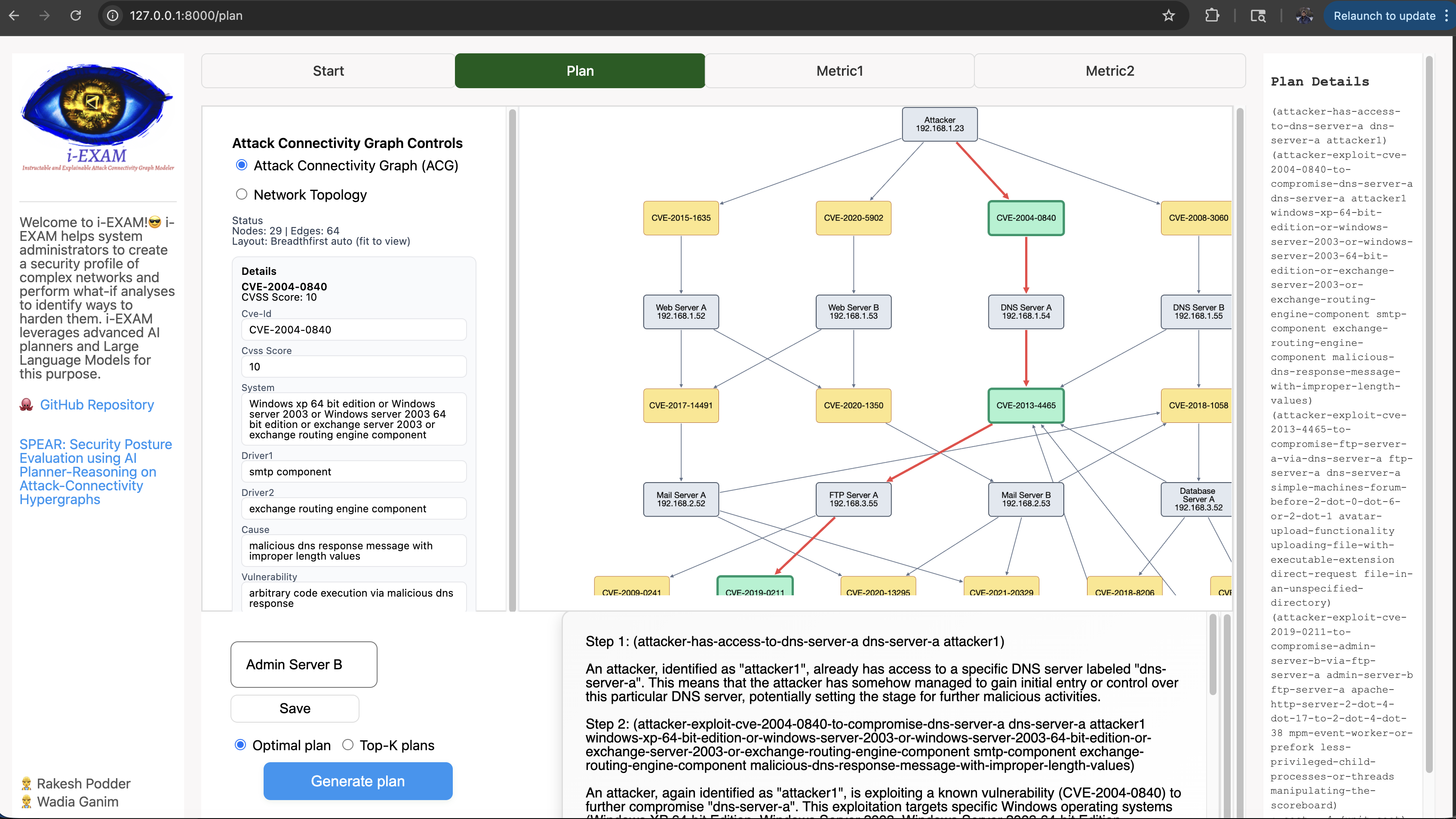}
    \caption{i-EXAM's Visual Interaction UI.}
    \label{fig:main-front}
\end{figure}

Fig. \ref{fig:main-front} is a screenshot from a case-study where i-EXAM was instantiated with a test network given the system configurations, network connectivity, CVEs, and a list of possible targets/goals. As shown in Fig. \ref{fig:main-front}, if the sysadmin selects a target to inspect (automatically highlighted in the visual graph), i-EXAM presents a plan to compromise the target. When the impenetrability metric is selected, 
i-EXAM outputs an attack plan and its cost, then prompts to explore proposed changes that would eliminate  attack paths for the target. i-EXAM presents diverse modification options to the sysadmin; upon selecting a modification option, the tool produces the corresponding solution and an explanation as to why no attack plan exists under that option. Similarly, under attack difficulty metric, i-EXAM prompts the sysadmin to select attacker initial positions; upon selection, it presents the optimal plan and cost, then prompts to explore proposed changes that would harden the network (by increasing cost of the attack plan). 
Once an option is selected, the tool applies the modification, recomputes the minimum-cost attack plan, and presents the results with an explanation of why the cost of attack has increased.

\vspace{-0.1in}


\section{Discussion}

{\bf Robustness.} i-EXAM inherits SPEAR's soundness and completeness guarantees (\cite{podder2025spear}, Theorem 1) such that any valid plan in the model corresponds to a valid attack or connectivity path in the ACG, and vice versa. A* search with the LMCut heuristic guarantees optimal hardening solutions when costs are specified. 
Diverse solution sets further guard against impracticality — if one hardening strategy cannot be implemented, alternatives remain.\\
{\bf Generalization.} The domain independent nature of PDDL-based ACG formalism generalizes across arbitrary network topologies and vulnerability combinations. The automated pipeline (Nmap, Wazuh, OpenVAS $\rightarrow$ Structure JSON $\rightarrow$ PDDL) constructs planning models for any network without manual modeling effort. The LLM explanation component generalizes across different plan-failure scenarios without task-specific fine-tuning, translating structured counterfactual plan traces into natural language for any network.\\
{\bf Scalability.} i-EXAM uses FastDownward with LMCut for plan computation and a modified A* search over model space for hardening. Empirical evaluation on networks of 30 nodes \cite{podder2025spear} shows that heuristic guidance consistently reduces computation time ($\sim$50\%) across both impenetrability and attack difficulty metrics. The top-k planner enables enumeration of diverse attack paths even in larger networks. \\
{\bf Deployment Effort.} Sysadmins require no knowledge of PDDL or AI planning. i-EXAM automates model construction from standard network scan outputs. The only inputs required are network access for scanning and a specification of target nodes and security goals. The human-in-the-loop design ensures administrators can apply domain judgment when selecting among diverse hardening options.




 \section{Conclusion \& Future Work}

i-EXAM bridges formal AI planning theory and real-world network security, demonstrating how PDDL-based reasoning and LLMs can be combined to make planning-based security analysis accessible to sysadmins. Future work includes learned cost functions for hardening actions, usability studies with domain experts, and extension to larger enterprise networks.

\section*{Acknowledgment}
This work was partially supported by the U.S.\ Office of Naval Research under award N000142612041.  The opinions, results, conclusions, and suggestions presented in this work are those of the authors and do not necessarily represent those of the Office of Naval Research or other organizations and agencies.
\bibliography{aaai2026}


\end{document}